# Maximum Entropy estimation of probability distribution of variables in higher dimensions from lower dimensional data

Jayajit Das[1-4], Sayak Mukherjee[1,2], and, Susan E. Hodge[1,2]

[1]Battelle Center for Mathematical Medicine, Research Institute at the Nationwide Children's Hospital, 700 Children's Drive, OH 43205

Departments of [2]Pediatrics, [3]Physics, and the [4]Biophysics Program, the Ohio State University, Columbus, OH

**Abstract**

A common statistical situation concerns inferring an unknown distribution Q(x) from a known distribution P(y), where X (dimension n), and Y (dimension m) have a known functional relationship. Most commonly, n≤m, and the task is relatively straightforward. For example, if $Y_1$ and $Y_2$ are independent random variables, each uniform on [0, 1], one can determine the distribution of $X = Y_1 + Y_2$; here m=2 and n=1. However, biological and physical situations can arise where n>m. In general, in the absence of additional information, there is no unique solution to Q in those cases. Nevertheless, one may still want to draw some inferences about Q. To this end, we propose a novel maximum entropy (MaxEnt) approach that estimates Q(x) based only on the available data, namely, P(y). The method has the additional advantage that one does not need to explicitly calculate the Lagrange multipliers. In this paper we develop the approach, for both discrete and continuous probability distributions, and demonstrate its validity. We give an intuitive justification as well, and we illustrate with examples.

**Introduction**

We are often interested in quantitative details about quantities that are difficult or even impossible to measure directly. In many cases we may be fortunate enough to find measureable quantities that are related to our variables of interest. Such examples are abundant in nature. Consider a community of microbes coexisting in humans or other metazoan species(1, 2). It is possible to measure the relative abundances of different species in the microbial community in individual hosts, but it could be difficult to directly measure parameters that regulate interspecies interactions in these diverse communities. Knowing the quantitative values of the parameters representing microbial interactions is of great interest, both because of their role in development of therapeutic strategies against diseases such as colitis, and for basic understanding, as we have discussed in (3).

Inference of these unknown variables from the available data is a subject of a vast literature in diverse disciplines including statistics, information theory, and, machine learning(4-7). In this paper we will be interested in a specific problem where the unknown variables in a large dimension are related to a smaller number of variables whose joint probability distribution is known from measurements.

In the above example, parameters describing microbial interactions could represent such unknown variables, and their number could be substantially larger than the number of measurable variables, such as abundances of distinct microbial species. The distribution of abundances of microbial species in a host population can be calculated from measurements performed on a large number of individual subjects. The challenge is to estimate the distribution of microbial interaction parameters using the distribution of microbial abundances.

These inference problems can be dealt with by Maximum Entropy(MaxEnt)-based methods that maximize an entropy function subject to constraints provided by the expectation values calculated from measured data(4, 5, 7, 8). In standard applications of MaxEnt, usually, averages, covariances, and, sometimes, higher-order moments calculated from the data are used to infer such distributions(4, 5, 7). Including larger number of constraints in the MaxEnt formalism involves calculating a large number of Lagrange multipliers by solving an equal number of nonlinear equations, which can pose a great computational challenge(9). Here we propose a novel MaxEnt-based method to infer the distribution of the unknown variables. Our method uses the distribution of the measured variables and provides an elegant MaxEnt solution that bypasses direct calculation of the Lagrange multiplers. Instead, the inferred distribution is described in terms of a degeneracy factor, described by a closed form expression, which depends only on the symmetry properties of the relation between the measured and the unknown variables.

More generally, the above problem relates to the issue of calculating a probability function of *X* from the probability function of *Y*, where *X* and *Y* are both random variables, and *Y* and X have a functional relationship. This could involve either discrete or continuous random variables. Standard textbooks (10) in probability theory usually deal with cases where X resides in a manifold (dimension *m*) of lower or the same dimension as the Y manifold (dimension *n*) and the distribution of the Y variables (y) at the higher dimension is known. We address the problem when n>m, where we estimate Q(x) from P(y). I.e., we infer the higher-dimension variable from the lower-dimension one. We show that when the variables are discrete, no unique solution exists for Q(x), as the system is underdetermined. However, the MaxEnt-based method can provide a MaxEnt solution in this situation that is constrained only by the available information (P(y) in this case) and is free from any additional assumptions. We then extend the results for continuous variables.

**The problem**

We state the problem, illustrating in this section with discrete random variables. Consider a case when n different random variables, $x_1,..,x_n$, are related to m (n>m) different variables, $y_1,..,y_m$, as $\{Y_i=f_i(x_1,..,x_n)\}$ (f: $R^n \to R^m$). We know the probabilities for the y variables and want to reach some conclusion about the probabilities of the x variables.

We introduce a few terms and notations borrowed from physics that we will use to simplify the mathematical description(11). A state in the x (or y) space refers to a particular set of values in the variables $x_1,..,x_n$ (or $y_1,..,y_m$). We denote the set of these states as $\{x_1,..,x_n\}$ or $\{y_1,..,y_m\}$. The vector notations, $\vec{x} = (x_1,\cdots,x_n)$ and $\vec{y} = (y_1,\cdots,y_m)$, will be used to compactly describe expressions when required. For the same reason, when we use f without a subscript, it will refer to a vector of f values, i.e., $\vec{y} = f(\vec{x}) = (f_1(\vec{x}),\ldots,f_m(\vec{x}))$. In standard textbook examples in elementary probability theory and physics, we are provided with the probability distribution function $P(\vec{y})$, where y is of higher or equal dimension (m≥n). Then $Q(\vec{x})$, with lower dimension n, is calculated using

$$Q(\vec{x}) = \sum_{y_1,\cdots,y_m} P(\vec{y}) \qquad (1a)$$

The summation in Eq. (1a) is performed over only those states $\{y_1,..,y_m\}$ that correspond to the specified state $\vec{x}$.

Here we are interested in the inverse problem: we are still provided with the probability distribution $P(y_1,..,y_m)$ and need to estimate the probability distribution $Q(x_1,..,x_n)$, but now m < n. No unique solution for $Q(x_1,..,x_n)$ exists in this situation as the system is underdetermined. Instead of (1a) we use this equation:

$$P(\vec{y}) = \sum_{x_1,\cdots,x_n | f(\vec{x}) = \vec{y}} Q(\vec{x}) = \sum_{x_1,\cdots,x_n} Q(\vec{x}) \prod_{i=1}^{m} \delta_{y_i, f_i(\vec{x})} \qquad (1b)$$

The constraints imposed on the summation in the last term by the relations ($\vec{y} = f(\vec{x})$) between the states in x and y are incorporated using the Kronecker delta function ($\delta_{ab}$, where, $\delta_{a,b}=1$ when a=b, and, $\delta_{a,b}=0$ when a≠b). For pedagogical reasons we elucidate the problem of non-uniqueness in the solutions using a simple example. This example can be easily generalized.

**Example I.** We start with a discrete random variable y, with known distribution $P(y) = 1/3$ for $y = 0, 1, 2$. Then assume that discrete random variables $x_1$ and $x_2$ are related to y, as, $y=f(x_1,x_2)=x_1+x_2$. We restrict $x_1$ and $x_2$ to being nonnegative integers; hence $x_1$, and $x_2$ can assume only three values, 0,1, and 2.

It follows that $Q(x_1,x_2)$ are related to $P(y)$ following Eq. (1b) as,

$$P(y) = \begin{cases} Q(0,0) \text{ for } y=0 \\ Q(0,1)+Q(1,0) \text{ for } y=1 \\ Q(0,2)+Q(1,1)+Q(2,0) \text{ for } y=2 \end{cases}$$

Hence

$$Q(0,0) = 1/3$$
$$Q(0,1)+Q(1,0) = 1/3$$
$$Q(0,2)+Q(1,1)+Q(2,0) = 1/3 \quad (2)$$

The above relation provides three independent linear equations for determining six unknown variables, $Q(0,0)$, $Q(1,0)$, $Q(0,1)$, $Q(1,1)$, $Q(2,0)$, and, $Q(0,2)$. Note, the condition of

$\sum_{x_1,x_2} Q(x_1,x_2) = \sum_y P(y) = 1$ is satisfied by the above linear equations, which also makes

$Q(1,2)=Q(2,1)=Q(2,2)=0$. Therefore, the linear system in Eq. (2) is underdetermined and $Q(x_1,x_2)$ cannot be found uniquely using these equations. (E.g., $Q(0,1)$ and $Q(1,0)$ could each equal 1/6; or $Q(0,1)$ could equal 1/12, with $Q(1,0)=1/4$; etc.)

This issue of non-uniqueness is general and will hold as long as the number of constraints imposed by $P(y_1,..,y_m)$ is smaller than that of the number of unknown $Q(x_1,..,x_n)$. For example, when each direction in y (or x) can take L (or $L_1$) discrete values and all the states in x are mapped to all the states in y, then the system will be underdetermined as long as, $L^m < L_1^n$.

**A MaxEnt based solution (discrete)**

In this section we propose a solution of this problem using a Maximum Entropy based principle, for discrete variables. We can define Shannon's entropy(4, 5, 7), S, given by

$$S = -\sum_{x_1,\cdots,x_n} Q(\vec{x}) \ln Q(\vec{x}) \tag{3}$$

and then maximize S with the constraint that $Q(\vec{x})$ should generate the distribution $P(\vec{y})$ in Eq. (1b).

Eq. (1b) describes the set of constraints spanning the distinct states in the y space. For example, when each element in the y vector assumes binary values (+1 or -1) there are in total $2^m$ number of distinct states in the y space providing $2^m$ number of equations of constraints. We can introduce a Lagrange multiplier for each of the constraint equations, which we denote compactly as a function, $\lambda(y_1,\ldots,y_m)$ or $\lambda(\vec{y})$ describing a map from $R^n \to R$. That is, every possible y vector is associated with a unique value of $\lambda$. Also note, when $P(\vec{y})$ is normalized, $Q(\vec{x})$ is normalized due to Eq. (1b), therefore, we will not use any additional Lagrange multiplier for the normalization condition of $Q(\vec{x})$.

We use the Lagrange multipliers(4, 11) to find the solution for $Q(\vec{x})$ that maximizes S:

$$\hat{Q}(\vec{x}) = Z^{-1} e^{-\sum_{y_1,\cdots,y_m} \lambda(\vec{y}) \prod_{i=1}^{m} \delta_{y_i, f_i(\vec{x})}} = Z^{-1} e^{-\lambda(f(\vec{x}))} \tag{4}$$

The partition sum, Z, is defined as,

$$1 = \sum_{x_1,\cdots,x_n} Q(\vec{x}) = Z^{-1} \sum_{x_1,\cdots,x_n} e^{-\lambda(f(\vec{x}))}$$
$$\Rightarrow Z = \sum_{x_1,\cdots,x_n} e^{-\lambda(f(\vec{x}))} \tag{5}$$

From the above solution (Eqs. (4) and (5)) we immediately observe the two main features that $Q(\vec{x})$ exhibits:

(i) The values of $\hat{Q}(\vec{x})$ for the states $\{x_1,\ldots,x_n\}$ that map to the same state $y_1,\ldots,y_m$ via $\{f_i(\vec{x})\}$ are equal to each other. In the simple example above, this implies Q(1,0)=Q(0,1), and, Q(1,1)=Q(0,2)=Q(2,0).

(ii) $\hat{Q}(\vec{x})$ contains all the symmetry properties present in the relation $\{y_i=f_i(x_1, \ldots, x_n)\}$. In the simple example, the relation between y and x was symmetric in permutation of $x_1$ and $x_2$, implying, $Q(x_1,x_2)=Q(x_2,x_1)$.

We will take advantage of the above properties to avoid direct calculation of the Lagrange multipliers in Eq. (4): For the states $\{\tilde{x}_1,\cdots,\tilde{x}_n\}$ in the x space that map to the same state, $\tilde{y}_1,\cdots,\tilde{y}_m$, in the y space, Eq. (1b) can rewritten as

$$P(\vec{\tilde{y}}) = \sum_{\tilde{x}_1,\cdots,\tilde{x}_n} \hat{Q}(\vec{\tilde{x}}) = k(\vec{\tilde{y}})\hat{Q}(\vec{\tilde{x}'}) \qquad (6a)$$

$$\Rightarrow \hat{Q}(\vec{\tilde{x}'}) = P(\vec{\tilde{y}})/k(\vec{\tilde{y}}) \qquad (6b)$$

where $k(\tilde{y}_1,\cdots,\tilde{y}_m)$ gives the total number of distinct states $\{\tilde{x}_1,\cdots,\tilde{x}_n\}$ in the x space that correspond to the state, $\tilde{y}_1,\cdots,\tilde{y}_m$ or $\vec{\tilde{y}}$. Since, all the states in $\{\tilde{x}_1,\cdots,\tilde{x}_n\}$ will have the same probability, in the second step in Eq. (6a) we replace the summation with $k(\vec{\tilde{y}})$, multiplied by the probability of any state $(\tilde{x}'_1,\cdots,\tilde{x}'_n)$ or $\vec{\tilde{x}'}$ in $\{\tilde{x}_1,\cdots,\tilde{x}_n\}$. We designate $k(\vec{\tilde{y}})$ as the degeneracy factor, borrowing a similar terminology in physics. $k(\vec{\tilde{y}})$ can be expressed in terms of the Kronecker delta functions as,

$$k(\vec{\tilde{y}}) = \sum_{x_1,\cdots,x_n} \left[ \prod_{i=1}^{m} \delta_{\tilde{y}_i, f_i(\vec{x})} \right] \qquad (7)$$

Note, the degeneracy factor in Eq. (7) only depends on the relationship between $\{\vec{x}\}$ and $\{\vec{y}\}$, and, does not depend on the probability distributions, P and Q. In our simple example above, since $y=x_1+x_2$, $Q(0,1)$ and $Q(1,0)$ both correspond to $y = 1$, therefore $k(\tilde{y}=1)=2$. Eq. (6b) is the main result of this section, which describes the inferred distribution $\hat{Q}(\vec{x})$ in terms of the known probability distribution $P(\vec{y})$, and, $k(\vec{y})$, which can be calculated from the given relation between y and x. Thus, the calculation of $\hat{Q}(\vec{x})$, as shown in Eq. (6b), does not involve direct evaluation of the Lagrange multipliers, $\lambda(\vec{y})$, or the partition sum, Z. These two quantities are related to $P(\vec{y})$, and, $k(\vec{y})$, following Eqs. (4), (6b) and (7), as,

$$Z^{-1}e^{-\lambda(f(\vec{x}))} = \frac{P(\vec{\tilde{y}})}{k(\vec{\tilde{y}})} \qquad (8)$$

Notice, $\lambda(\vec{\tilde{y}})$, and, $Z$, cannot be calculated uniquely from the above equation and their individual values do not affect $\hat{Q}(\vec{x})$, as long as, they are related through Eq. (8).

**Example I, continued.** We provide a solution for Example I presented above. By simple counting, we see the degeneracy factors are

$k(\tilde{y}=0)=1, k(\tilde{y}=1)=2, k(\tilde{y}=2)=3$

Thus following Eq. (2), Q(0,0)=P(0)=1/3, Q(0,1)=Q(1,0)=P(1)/2=1/6, and, Q(2,0)=Q(1,1)=Q(0,2)=P(2)/3=1/9. For more complex problems, the degeneracy factors can be calculated numerically. Maximizing the entropy, S, is what made all the Qs be equal for any one y value.

**Results for continuous variables**

The above results can be extended when $\{X_i\}$ and $\{Y_i\}$ are continuous variables. However, there is an issue that make a straightforward extension of the calculations shown in the discrete case in the continuum limit difficult. The issue is related to the continuum limit of the entropy function S in Eq. (3). Replacing the summation in Eq. (3) with an integral in the limit of large number of states as the step size separating the adjacent states is decreased to zero creates an entropy expression which is negative and unbounded(12). This problem can be ameliorated by defining a relative entropy, RE, defined as,

$$RE = \int dx_1 \cdots dx_n q(x_1,\cdots,x_n) \ln\left[\frac{q(x_1,\cdots,x_n)}{u(x_1,\cdots,x_n)}\right] \qquad (9)$$

where, u is a uniform pdf, defined on the same domain as q. RE always remains positive with a lower bound at zero. The results obtained by maximizing S in the previous section can be derived by minimizing a relative entropy defined with the discrete distributions, Q and a uniform distribution, U, where the integral in Eq. (9) is replaced by a summation over the states in the x space. RE in Eq. (9) quantifies the difference between the distribution $q(x_1,\ldots,x_n)$ and the corresponding uniform distribution.

The definition of RE in Eq. (9) still has an issue of defining the uniform distribution when the x variables are unbounded. In some cases, it may be possible to solve the problem by introducing finite upper and lower bounds and then analyzing the results in the limit where the upper (or lower) bound approaches ∞ (or -∞). We will illustrate this approach in Example IV, below. Also, see Example III for a comparison.

In the continuum limit, the constraints on $q(x_1,..,x_n)$ or $q(\vec{x})$, imposed by the p.d.f. $p(y_1,..,y_m)$ or $p(\vec{y})$ are given by,

$$p(\vec{y}) = \int dx_1 \cdots dx_n q(\vec{x}) \prod_{i=1}^{m} \delta_D(y_i - f_i(\vec{x})) \tag{10}$$

The Dirac delta function for a single variable x is defined as,

$$\int_R dx \, \delta_D(x) = 1 \tag{11}$$

where the region R contains the point x=0.

Since, the pdf $p(\vec{y})$ resides in a lower dimension compared to $q(\vec{x})$, estimation of $q(\vec{x})$ in terms of $p(\vec{y})$ requires solution of an underdetermined system.

For continuous variables we can proceed with minimizing the relative entropy using functional calculus(13, 14).

The relative entropy RE in Eq. (9) is a functional of $q(\vec{x})$. As in the discrete case, if $p(\vec{y})$ is normalized, i.e., $\int dy_1 \cdots dy_m p(\vec{y}) = 1$, then Eqs. (10) and (11) imply $q(\vec{x})$ is normalized as well, i.e.,

$$\int dx_1 \cdots dx_n q(\vec{x}) = 1 \tag{12}$$

We introduce a Lagrange multiplier function, $\lambda(\vec{y})$, and generate a functional, $S_\lambda[q]$, that we need to minimize in order to minimize Eq. (9) along with the constraints in Eq.(10). Since, the normalization condition in Eq. (12) follows from Eq. (10) we do not treat Eq. (12) as a separate constraint.

$S_\lambda[q]$ is given by,

$$\begin{aligned} S_\lambda[q(\vec{x})] &= \int dx_1 \cdots dx_n \left[ q(\vec{x}) \ln\left[\frac{q(\vec{x})}{u(\vec{x})}\right] \right] \\ &\quad - \int dy_1 \cdots dy_m \lambda(\vec{y}) \left[ P(\{y_j\}) - \int dx_1 \cdots dx_n q(\vec{x}) \prod_{k=1}^{m} \delta_D(y_k - f_k(\vec{x})) \right] \\ &= \int dx_1 \cdots dx_n \left[ q(\vec{x}) \ln\left[\frac{q(\vec{x})}{u(\vec{x})}\right] \right] - \int dy_1 \cdots dy_m \lambda(\vec{y}) p(\vec{y}) \\ &\quad - \int dx_1 \cdots dx_n \, \lambda(f(\vec{x})) q(\vec{x}) \prod_{k=1}^{m} \delta_D(y_k - f_k(\vec{x})) \end{aligned} \tag{13}$$

We can take the functional derivative to minimize S as,

$$\frac{\delta S[q(\vec{x})]}{\delta q(\vec{x})} = \ln[q(\vec{x})] + 1 - \ln[u(\vec{x})] - \lambda(f(\vec{x})) = 0 \tag{14}$$

In deriving Eq. (14) we used the standard relation $\frac{\delta f[x]}{\delta f[x']} = \delta_D(x - x')$. For multiple dimensions this generalizes to,

$$\frac{\delta f[\vec{x}]}{\delta f[\vec{x}']} = \prod_{j=1}^{n}\delta_D(x_j - x'_j)$$ The chain rule for derivatives of functions can be easily generalized for functional derivatives(14).

Eq. (14) provides us with the solution that minimizes Eq. (13):

$$\hat{q}(\vec{x}) = u(\vec{x})e^{\lambda(f(\vec{x}))+1} = u_0 e^{\lambda(f(\vec{x}))+1} = \hat{q}(f(\vec{x})) \quad (15)$$

where, $u_0$ is a constant related to the density of the uniform distribution.

Note the $\{x_i\}$ dependence in the solution, $\hat{q}(\vec{x})$, arises only though $f(\vec{x})$.

Substituting Eq. (15) in Eq. (10),

$$p(\vec{y}) = \int dx_1 \cdots dx_n \hat{q}(\vec{x}) \prod_{k=1}^{m}\delta_D(y_k - f_k(\vec{x}))$$

$$= \int dx_1 \cdots dx_n \hat{q}(f(\vec{x})) \prod_{k=1}^{m}\delta_D(y_k - f_k(\vec{x})) = \hat{q}(\vec{y})\kappa(\vec{y})$$

$$\Rightarrow \hat{q}(\vec{y}) = \frac{p(\vec{y})}{\kappa(\vec{y})} \quad (16)$$

where,

$$\kappa(\vec{y}) = \int dx_1 \cdots dx_n \prod_{k=1}^{m}\delta_D(y_k - f_k(\vec{x})) \quad (17)$$

The second derivative gives,

$$\frac{\delta^2 S[q(\vec{x})]}{\delta q(\vec{x}')\delta q(\vec{x}'')} = 1/q(\vec{x}'')\delta_D(x' - x'') \quad (18)$$

The second derivative of S in Eq. (18) is always positive, since q is positive. Therefore, $\hat{q}(\vec{x})$, minimizes the relative entropy in Eq. (9). Equations (16) and (17) are the main results of this section, which are the counterparts for the equations (6b) and (7) in discrete case.

We apply the above results for two examples below.

**Example II.**

Consider a linear relationship between y and x, e.g., y=$x_1$+$x_2$, where, 0≤y≤∞ and 0≤ $x_1$≤∞, 0≤ $x_2$≤∞. If the p.d.f in y is known as, p(y)=1/µ exp(-y/µ), we would like to know the p.d.f corresponding q($x_1$,$x_2$), where, the pdfs p and q are related by Eq. (10), i.e.,

$$p(y) = \int_0^\infty dx_1 \int_0^\infty dx_2 \, q(x_1,x_2) \delta_D(y - x_1 - x_2)$$

The degeneracy factor in the continuous case, according to Eq. (17), in this case is,

$$\kappa(y) = \int_0^\infty dx_1 \int_0^\infty dx_2 \, \delta_D(y - x_1 - x_2) = \int_0^y dx_1 \int_0^y dx_2 \, \delta_D(y - x_1 - x_2)$$

$$= \int_0^y dx_1 \int_0^y dx_2 \, \delta_D(x_2 - (y - x_1)) = \int_0^y dx_1 = y$$

The second equality results from the fact that the Dirac delta function is zero outside that region. The fourth equality uses the property of the Delta function,

$$\int_0^y dx_2 \delta_D(x_2 - a) \text{ (where, } a = y - x_1 = const \text{ for this integration)}$$
$$= 1$$

Therefore, $\hat{q}(x_1,x_2) = \dfrac{e^{-(x_1+x_2)/\mu}}{\mu(x_1 + x_2)}$ .

**Example III.**

Let $y = f(x_1,x_2) = x_1^2 + x_2^2$ , 0≤y≤∞ and 0≤ ($x_1$, $x_2$)≤∞ . Then κ(y), as given by Eq. (17), is,

$$\kappa(y) = \int_0^\infty dx_1 \int_0^\infty dx_2 \, \delta_D(y - x_1^2 - x_2^2) = \int_0^\infty dr\, r\, \delta_D(y - r^2) \int_0^{\pi/2} d\phi = \int_0^\infty d(r^2) \, \delta_D(y - r^2) \frac{\pi}{4}$$

$$= \frac{\pi}{4}$$

Therefore, according to Eq. (17),

$$\hat{q}(y) = \frac{4 p(y)}{\pi} \Rightarrow \hat{q}(x_1,x_2) = \frac{4 p(f(x_1,x_2))}{\pi}$$

In our final example, we illustrate solving the problem by taking the limit when the upper and/or lower bound(s) approach $\pm\infty$, as mentioned near the beginning of this section of the paper.

**Example IV.**

Let $y = x_1^2 + x_2^2$, $0 \leq y \leq 2L^2$ and $0 \leq (x_1, x_2) \leq L$. First we calculate $\kappa(y)$ as given in Eq. (17). Therefore, we need to evaluate the integral,

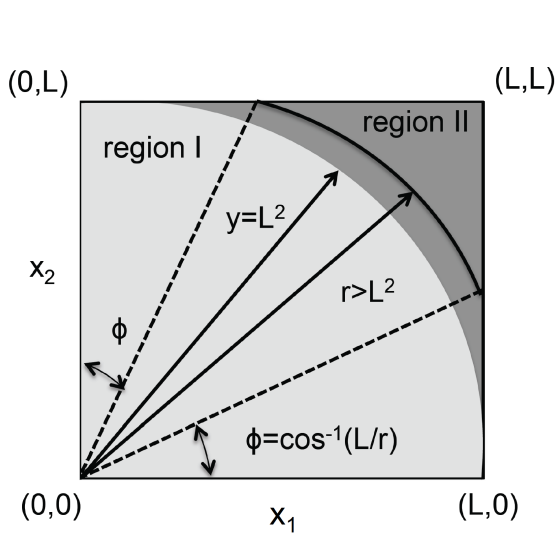

$$\kappa(y) = \int_0^L dx_1 \int_0^L dx_2 \delta_D(y - x_1^2 - x_2^2)$$

We divide the region of integration ($0 \leq (x_1, x_2) \leq L$) into two parts, region I (lighter shade) and II (darker shade) as shown in the figure. Region I contains $x_1$ and $x_2$ values, where, $x_1^2 + x_2^2 = y^2 \leq L^2$, and, region II contains the remaining of the part of the domain ($0 \leq (x_1, x_2) \leq L$) of integration. The integrals in these regions are given by the first and the second term after the second equality sign in the equation below.

$$\kappa(y) = \int\int_{region\ I} dx_1 dx_2 \delta_D(y - x_1^2 - x_2^2) + \int\int_{region\ II} dx_1 dx_2 \delta_D(y - x_1^2 - x_2^2)$$

$$= \int_0^L drr\, \delta_D(y - r^2) \int_0^{\pi/2} d\phi + \int_0^L drr \int_{\cos^{-1}(r/L)}^{\pi/2 - \cos^{-1}(r/L)} d\phi \delta_D(y - r^2)$$

In region I, where $y \leq L^2$,

$$\kappa_I(y) = \int_0^L drr\, \delta_D(y - r^2) \int_0^{\pi/2} d\phi = \frac{\pi}{2} \frac{\sqrt{y}}{2\sqrt{y}} = \frac{\pi}{4}$$

In region II, where $L^2 \leq y \leq 2L^2$,

$$\kappa_{II}(y) = \int_0^L dr\, r \int_{\cos^{-1}(L/r)}^{\pi/2 - \cos^{-1}(L/r)} d\phi\, \delta_D(y - r^2) = \int_0^L dr\, r (\pi/2 - 2\cos^{-1}(L/r))\delta_D(y - r^2)$$

$$= \frac{\sqrt{y}}{2\sqrt{y}}(\pi/2 - 2\cos^{-1}(L/\sqrt{y})) = \frac{\pi}{4} - \cos^{-1}(L/\sqrt{y})$$

$\cos^{-1}(L/\sqrt{y})$ varies between 0 (on the line $x_1^2 + x_2^2 = L^2$) and $\pi/4$ (at $x_1=x_2=L$). Note, $\kappa(y)=0$ when $x_1=x_2=L$, which does have any degeneracy. Therefore, Eq. (16) is not valid at this point. Thus, as in Example II and III,

$$\hat{q}(x_1, x_2) = \frac{4p(f(x_1, x_2))}{\pi}, \text{ when, } x_1^2 + x_2^2 \leq L^2$$

$$= \frac{p(f(x_1, x_2))}{\left(\pi/4 - \cos^{-1}\left(L/\sqrt{x_1^2 + x_2^2}\right)\right)} \text{when, } L^2 \leq x_1^2 + x_2^2 < 2L^2$$

Limit $L \to \infty$: When $y \leq L^2$, $\kappa(y)=\pi/4$. Thus, as $L \to \infty$, as long as, remains in region I we correctly recover the result in example III. If y is in region II, then we can expand $\kappa(y)$ in a series of a small parameter $\varepsilon=(y-L^2)/L^2$ as $\kappa(y)=\pi/4-\varepsilon/2+O(\varepsilon^3)$. This result follows from the expansion of $L/\sqrt{y} < 1$ in region II. We can write,

$$L/\sqrt{y} = 1/\sqrt{1+\varepsilon} = (1+\varepsilon)^{-1/2} = 1 - \varepsilon/2 + \varepsilon^2/4 - O(\varepsilon^3)$$

where, $0<\varepsilon[=(y-L^2)/L^2] \leq 1$. Using series expansion of $\cos^{-1}(x)$ (15) we find,

$$\cos^{-1}(L/\sqrt{y}) = \sin^{-1}(\sqrt{1-(L/\sqrt{y})^2}) = \sin^{-1}(\varepsilon/2 + O(\varepsilon^2)) = \varepsilon/2 + O(\varepsilon^3)$$

and thus, $\kappa(y)=\pi/4-\varepsilon/2+O(\varepsilon^3)$.

**Discussion**

The problem we have attacked here arose from our work with microbial communities, e.g., ref. 3), but it also has broader statistical applications. For example, the responses of immune cells to external stimuli involve protein interaction networks, where protein-protein interactions, described by biochemical reaction rates, are not directly accessible for measurement *in vivo*. Recent developments in single cell measurement techniques allow for measuring many protein abundances in single cells, making it possible to evaluate distribution of protein abundances in a cell population(16). However, it is a challenge to characterize protein-protein interactions underlying a cellular response because the number of these interactions could be substantially larger than the number of measured protein species(17). These problems involve determining the distribution of a random variable x, where y is another random variable, and X and Y have a functional relationship. In the more common situation, x has dimensionality less than or equal to that of y, and there is often a unique solution. In contrast, we considered here the case where x's dimensionality is greater than that of y, so there is no unique solution to the problem.

Since there is no unique solution, we propose taking a MaxEnt approach, as a way of "spreading out the uncertainty" as evenly as possible. In the discrete case, intuition would suggest that if *k* values of Q sum to a given value of P, then the solution that makes the least additional assumptions is for each Q to equal P/k. This intuition is confirmed by our MaxEnt results for the discrete case. In the continuous case, the intuition is not as obvious. However, the MaxEnt solution does capture the same intuitive idea. Instead of dividing P by $k(\vec{y})$ (an integer), we divide p by $\kappa(\vec{y})$, where $\kappa(y) = \int_{\vec{x}} d\vec{x}\, \delta_D(y - f(\vec{x}))$ when *y* has dimension 1, or more generally by eq. (16). This use of the Dirac delta function has the similar effect of spreading out the uncertainty evenly.

Estimating the distribution Q(x) does not require explicit calculation of the Lagrange multipliers and the partition sum . Rather, Q(x) is directly evaluated following Eq. (6b) (or Eq. (17) in the continuous case), using the measured P(y) (or p(y)), and, k(y) (or κ(y)), which depends only on the relationship y=f(x). In standard MaxEnt applications, where constraints are imposed by the average values and other moments of the data, inference of probability distributions requires evaluation of the Lagrange multipliers and the partition sum Z. This involves solving a set of nonlinear equations and the relation between the Z and the Lagrange multipliers, with the imposed constraints depending on the microscopic details of the models. Calculating these quantities, which is usually carried out numerically, can pose a technical challenge when the variables reside in large dimensions. In our case, we avoid these calculations and provide a solution for $Q(x)$ in terms of a closed analytical expression, which is general and thus applicable to any well-behaved example. A limitation is that calculation of the degeneracy factor $k(y)$ (or $\kappa(y)$ in the continuous) case can present a challenge in higher dimensions and for complicated relations between y and x. Monte Carlo sampling techniques and discretization schemes for Dirac delta functions(18) can be helpful in that regard.

Another important difference between the cases we investigated here and the standard application of MaxEnt is the non-uniqueness of the partition sum, Z, and the Lagrange multipliers. Usually both possess unique values when expectation values are used as constraints. In our case, the non-uniqueness appears to be related to a Gauge transformation, in the relation $y=f(x)$, similar to the Gauge invariance in the spin glass models(19). The appearance of Gauge invariances can occur when, as aptly described by Brown and Sethna(20), "a model has more detail than nature provides".

**Acknowledgements:**

The work is supported by a grant from NIGMS (1R01GM103612-01A1) to JD. JD is also partially supported by The Research Institute at the Nationwide Children's Hospital and a grant from the Ohio Supercomputer Center (OSC). SEH is supported by The Research Institute at the Nationwide Children's Hospital. JD and SM thank Aleya Dhanji for carrying out preliminary calculations related to the project.

**Author contributions:** JD, SM, and, SEH planned the research, carried out the calculations, and, wrote the paper.

**Conflicts of interest:** Authors declare no conflicts of interest.
The founding sponsors had no role in the design of the study; in the collection, analyses, or interpretation of data; in the writing of the manuscript, and in the decision to publish the

results